\documentclass[traditabstract, letter]{aa}
\usepackage{txfonts}
\usepackage{graphicx}
\usepackage{float}
\usepackage{natbib}
\usepackage{hyperref}
\usepackage[english]{babel}

\begin{document}

\title{Where stars form and live at high redshift: clues from the infrared}

\author{M. B\'ethermin\inst{1,2} \and O. Dor\'e\inst{3,4} \and G. Lagache\inst{2}}

\institute{
\inst{1} Laboratoire AIM-Paris-Saclay, CEA/DSM/Irfu, CNRS, Univ. Paris Diderot, Saclay, France, Email: \url{matthieu.bethermin@cea.fr}\\
\inst{2} Institut d'Astrophysique Spatiale (IAS), bat121, F-91405 Orsay, France; Universit\'e Paris-Sud 11 and CNRS (UMR8617)\\
\inst{3} Jet Propulsion Laboratory, California Institute of Technology, 4800 Oak Grove Drive, Pasadena, California, U.S.A.\\
\inst{4} California Institute of Technology, Pasadena, California, U.S.A.\\
}

\date{Received 7 December 2011 / Accepted 24 December 2011}

\titlerunning{Where stars form and live at high redshift: clues from the infrared}

\authorrunning{B\'ethermin et al.}

\abstract{The relation between dark matter halos and the loci of star formation at high redshift is a pressing question in contemporary cosmology. Matching the abundance of halos to the abundance of infrared (IR) galaxies, we explore the link between dark matter halo mass ($M_h$), stellar mass ($M_\star$) and star-formation rate (SFR) up to a redshift of 2. Our findings are five-fold. First, we find a strong evolution of the relation between $M_\star$ and SFR as a function of redshift with an increase of $sSFR = SFR/M_\star$ by a factor $\sim$30 between z=0 and z= 2.3. Second, we observe a decrease of sSFR with stellar mass. These results reproduce observed trends at redshift z$>$0.3. Third, we find that the star formation is most efficient in dark matter halos with $M_h\simeq 5\times$10$^{11}$\,$M_\odot$, with hints of an increase of this mass with redshift. Fourth, we find that $SFR/M_h$ increases by a factor $\sim$15 between $z=0$ and $z=2.3$. Finally we find that the SFR density is dominated by halo masses close to $\sim$7$\times$10$^{11}$ M$_{\odot}$ at all redshift, with a rapid decrease at lower and higher halo masses. Despite its simplicity, our novel use of IR observations unveils some characteristic mass-scales governing star formation  at high redshift.}

\maketitle

\section{Introduction}

The loci of star formation at high redshift is one of the salient questions of galaxy formation theory. In particular, uncertainties in the relation between the host halo mass and the IR luminosity function (LFIR) currently limit models of the clustering of star-forming galaxies as revealed e.g. by the anisotropies of the cosmic infrared background (CIB). This comes from the fact that the clustering of galaxies is directly driven by the mass of their host halos. The so-called abundance matching technique is another valuable tool that provides insights on this relation \citep{Vale2004}. 
While unexplored in the IR, a recent improvement of LFIR measurements makes this approach now both promising and timely. We will illustrate its power in this letter. In particular, working in the IR provides the unique possibility to constrain the star-formation rate (SFR) directly without any extra assumption regarding the physical processes that drive the star formation. 
\\

Considering the latest observational and theoretical developments in the IR, we introduce in Sect.~\ref{AM_principle} the abundance matching technique. We describe in detail the mass and luminosity functions used in Sect.~\ref{MLF}. Matching abundances allows us to infer relations between halo and stellar mass (Sect.~\ref{sm_sh}), SFR and stellar mass (Sect.~\ref{sfr_sm}) and SFR and halo mass (Sect.~\ref{sfr_sh}). In these sections, we will illustrate in particular how this simple but well-informed approach satisfyingly reproduces trends observed in the IR, but also at other wavelengths. Focusing on what is required for the modeling of the CIB, we exclude the so-called {\it starbursting} galaxies here because they are non-significant outliers. Critical in the  context of abundance matching, this assumption is well motivated and discussed in Sect.~\ref{MLF}.  
\\
Throughout this paper, we adopt a \citet{Chabrier2003} initial mass function (IMF)\footnote{Stellar masses and SFR computed assuming a \citet{Salpeter1955} IMF can be converted to \citet{Chabrier2003} IMF applying a -0.24~dex correction.} and a $\Lambda$CDM WMAP-7 cosmology \citep{Larson2011}.

\section{\label{AM_principle}Simple abundance matching}

Our goal is to find an observationally supported relation between the SFR of galaxies, their stellar and their host halo mass. While this relation is statistical in nature, the abundance matching technique \citep[e.g.][]{Conroy2009} provides a way to relate the means of this relation.
Observations suggest a small scatter around this relation and we will therefore neglect it: 0.16~dex for the relationship between stellar and halo mass \citep{More2009}, and 0.2~dex between the stellar mass  and the SFR (\citet{Rodighiero2011}; see also \citet{Behroozi2010} for a more general discussion of scatter).

The key assumption of this matching technique is the existence of a continuous and monotonic relation between the halo mass, the stellar mass, and the SFR:
$SFR = f(M_\star)$,  $SFR = g(M_h)$  and $M_\star = h(M_h)$, 
where $SFR$ is the SFR, $M_\star$ the galaxy stellar mass (SM), and $M_h$ the dark-matter halo mass (HM). 

We solve for the functions $f$, $g$ and $h$ that satisfy the following relations between the infrared luminosity function, the halo mass function, and the stellar mass function:
\begin{eqnarray}
n_{L_{IR}} \left ( > L_{IR} = K^{-1} \times f \left (M_\star \right ) \right ) &=& n_{M_\star}(> M_\star)\\
n_{L_{IR}} \left ( > L_{IR} = K^{-1}  \times g \left (M_h \right ) \right ) &=& n_{M_h}(> M_h) \\
n_{M_\star}(> M_\star = h(M_h)) &=& n_{M_h}(> M_h), 
\end{eqnarray}
where $n_{L_{IR}}$ is the number of galaxies per unit comoving volume brighter than the bolometric infrared luminosity $L_{IR}$, and $n_{M_\star}$ and $n_{M_{h}}$ are the number of galaxies more massive than the stellar mass $M_\star$ and the halo mass $M_h$, respectively\footnote{Note that $n$ can be deduced from the luminosity/mass function: $n_X(>X_{min}) = \int_{X_{min}}^{+\infty} \frac{d^2N}{dXdV} \, dX$, where X is $L_{IR}$, $M_\star$, or $M_h$.}. $K$ is the \citet{Kennicutt1998} factor, which links the infrared bolometric luminosity integrated between 8 and 1000~$\mu$m and the SFR: $K= 10^{-10} M_\odot\, yr^{-1} L_\odot^{-1}$ \footnote{The Kennicutt factor was re-scaled from the Salpeter to the Chabrier IMF.}. As we discuss below, a key assumption in this matching is that all (sub)halos contribute to the IR.

\section{\label{MLF}Mass and luminosity functions}
The key inputs to our discussions are the bolometric infrared luminosity function (LFIR), the stellar mass function (SMF) and the halo mass function (HMF).

It is well known that the LFIR is inconsistent with the \citet{Schechter1976} form and presents an excess at the bright end \citep[e.g.][]{Saunders1990}. However, it has been shown that this excess is mainly caused by starbursting galaxies, which are well outside the SFR-SM main sequence. \citet{Yun2001} showed that the 60\,$\mu$m LF of the IRAS 2\,Jy sample can be fitted by the sum of two Schechter functions. More recently, \citet{Sargent2011} showed that the LFIR can be decomposed as a sum of two approximate Schechter functions. The first one is responsible for $\sim$85\% of the IR energy output, and is caused by the so-called {\it main-sequence} objects that have very similar specific sSFR at a given $M_\star$ and redshift. The second one dominates the bright-end of the LF, and is caused by starbursting galaxies with an sSSFR about 0.5~dex higher than the main sequence. These starbursts contribute only $\sim$2\% to the mass-selected star-forming galaxies number density \citep{Rodighiero2011}, and $\sim$15\% to the SFR density \citep{Sargent2011}, and consequently to the CIB. The former guarantees that starbursts constitute a negligible fraction of IR objects, i.e. do not contribute significantly to the IR abundances. The latter suggests that they can be neglected for our purpose. 
Additionally, because a bimodal relation between LFIR and $sSFR$ would violate our hypothesis of a monotonic relation between masses and SFRs, we henceforth ignore the starburst components. Focusing on the main-sequence population, we therefore write the LFIR as a \citet{Schechter1976} function
\begin{equation}
\frac{dN}{dL_{IR}} =  \phi_\star(z) \left ( \frac{L_{IR}}{L_\star(z)} \right )^{-\alpha} \textrm{exp} \left ( \frac{L_{IR}}{L_\star(z)} \right ) \frac{1}{L_\star(z)},
\end{equation}
where $dN/dL_{IR}$ is the LFIR, and $\phi_\star(z)$ and $L_\star(z)$ are the characteristic density and luminosity at a given redshift, respectively. We use the evolution of $\phi_\star(z)$ and $L_\star(z)$ measured by \citet{Magnelli2011} up to z=2.3.

In contrast to the LFIR, the SMF is much closer to a \citet{Schechter1976} function. \citet{Perez-Gonzalez2008} have measured the SMF up to z=4 and we used their parameters of the Schechter function fit. We performed our abundance matching at the same redshift as the center of the redshift bins of \citet{Perez-Gonzalez2008}. We checked that there is no inconsistency between the SFR derived from our LFIR and that derived from the SMF following \citet{Wilkins2008}.

For the HMF, we used the \citet{Tinker2008} $z$-dependent HMF and the associated sub-halo mass function proposed in \citet{Tinker2010} with $\Delta=200$.  Our results are fairly insensitive to the inclusion of the sub-halo mass function.

\section{Results}

\subsection{Relation between stellar mass and halo mass}
\label{sm_sh}
\begin{figure}[!h]
\centering
\includegraphics{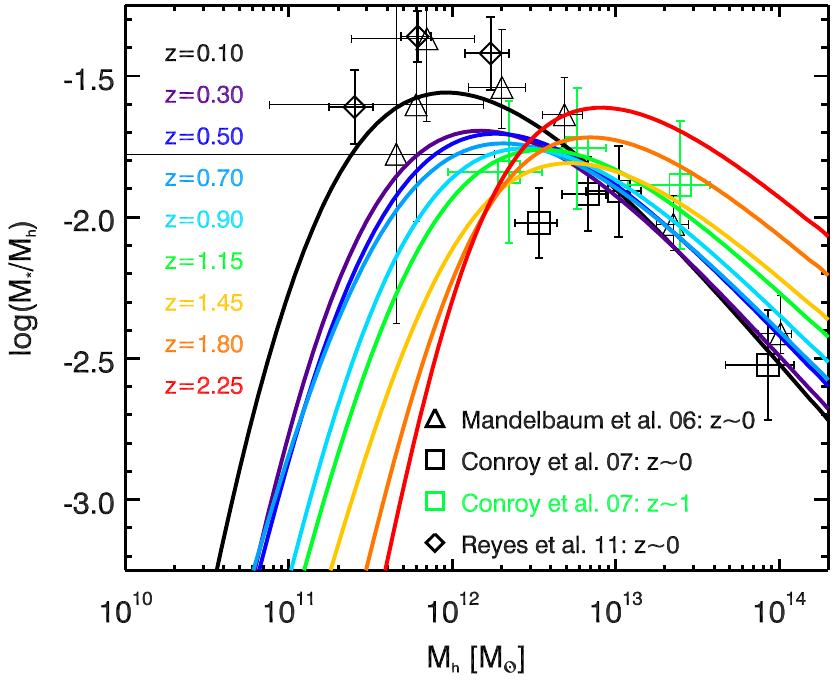}
\caption{\label{fig:starhalo} Ratio between stellar and halo mass as a function of halo mass from z=0 to z=2.25. \textit{Solid lines}: our results based on abundance matching technique. \textit{Triangles}: measurements from galaxy-galaxy lensing \citep{Mandelbaum2006}. \textit{Squares}: measurements from satellite galaxies \citep{Conroy2007} at z$\sim$0 (black) and z$\sim$1 (green). \textit{Diamonds}: measurements from combined kinematics and galaxy-galaxy lensing \citep{Reyes2011}. Typical uncertainties on the results of abundance matching are 0.2\,dex.}
\end{figure}

The abundance matching between the HMF and SMF has been extensively studied for z$<$1 (e.g. \citealt{Conroy2009}) and has been discussed in detail up to z$\sim$4 by \cite{Behroozi2010}. Because $M_\star/M_h$ as a function of $M_h$ is one basic element of our analysis, we first investigated it using our mass functions. Fig. \ref{fig:starhalo} shows this relation relation up to z$\sim$2.25. It agrees with the measurements at z$\sim$0 and z$\sim$1 \citep{Mandelbaum2006,Conroy2007,Reyes2011}. However, constraints are poor except at $z=0$ for $M_h>10^{11} \, M_\odot$. We find that $M_\star/M_h$ peaks near $M_h$=10$^{12}$~M$_\odot$ and decreases at lower- and higher- mass. This suggests that the integrated star-formation efficiency ($\equiv M_\star/M_h$) decreases for smaller as well as higher-mass halos. Fig. \ref{fig:starhalo} also indicates that the mass corresponding to the peak of integrated star-formation efficiency evolves with redshift, being a factor of 8 larger at z=2.25 than at z=0. We extended our analysis to higher redshifts using the latest SMF measurements \citep{caputi2011,Kajisawa2009} and confirm this trend. However, as discussed in \cite{Behroozi2010}, these results have to be taken with care because the current uncertainties make these evolutions quite uncertain.

\subsection{Relation between star-formation rate and stellar mass}
\label{sfr_sm}
 
\begin{figure}[!h]
\centering
\includegraphics{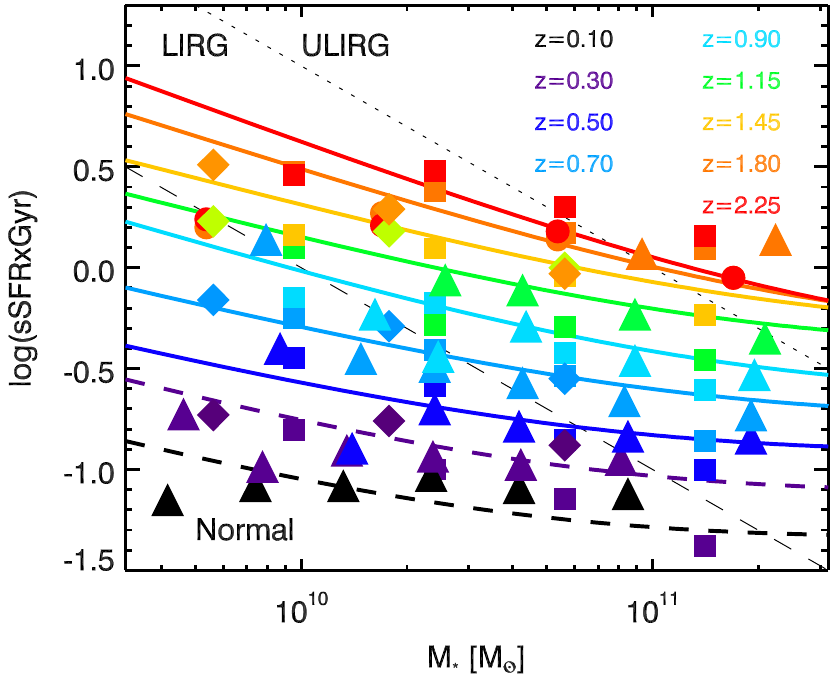}
\caption{\label{fig:ssfrstar} sSFR as a function of stellar mass at various redshifts. \textit{Solid lines}: our results based on the abundance matching technique (we use \textit{dashed lines} at $z \le 0.3$ where neglecting the quiescent galaxies have a significant impact, see Sect.~\ref{discussion}). \textit{Filled squares}: measurements from \citet{Karim2011}  obtained by radio stacking at $z=$0.3, 0.5, 0.7, 0.9, 1.1, 1.4, 1.8, and 2.25. \textit{Filled diamonds}: measurement from \citet{Rodighiero2010}  obtained by \textit{Herschel}/PACS stacking at $z=$0.25, 0.75, 1.25, and 1.75. \textit{Filled circles}: measurements from \citet{Bauer2011} obtained using UV measurements at $z=$1.75 and 2.25. \textit{Filled triangles}: measurements of \citet{Oliver2010} from \textit{Spitzer}/MIPS stacking at $z=$0.1, 0.3, 0.5, 0.7, 0.9, 1.1, 1.35, 1.75. The error bars have not been plotted for clarity, but they roughly match the symbol size. \textit{Thin-dashed} and \textit{dotted} lines are the limits between the normal and LIRG (luminous infrared galaxies, $10^{11} \, L_\odot< L_{IR} < 10^{12} \, L_\odot$), and the LIRG and ULIRG (ultra luminous infrared galaxies, $L_{IR} > 10^{12} \, L_\odot$) regime, respectively.}
\end{figure}

Matching the abundances derived from the LFIR and the SMF, we plot in Fig.~\ref{fig:ssfrstar} the sSFR as a function of $M_\star$ and redshift. We observe a strong increase of the sSFR with redshift (about 1.5 order of magnitude between $z=0$ and $z=2$), which qualitatively agrees with observations \citep{Elbaz2011,Karim2011}. We also see a slight decrease of the sSFR with $M_\star$ (0.5\,dex between $M_\star=10^{9.5} M_\odot$ and $M_\star=10^{11.5} M_\odot$), which again agrees with the observations \citep{Rodighiero2011,Karim2011}. To further illustrate this qualitative agreement, we also plot in Fig.~\ref{fig:ssfrstar} the most recent sSFR measurements. We reproduce their trends quite well. We note the large dispersion between the measurements, however. For example, \citet{Karim2011} and \citet{Rodighiero2011} differ significantly at high-mass at $z\sim$0.3, and at all masses for z$\sim$1 (note however that the redshift bins are quite different). Also, at higher redshift the measurements seem to differ by factors $\sim$2 around 10$^{10}$ M$_{\odot}$. 

\subsection{Relation between star-formation rate and halo mass}
\label{sfr_sh}
\begin{figure}[!h]
\centering
\includegraphics{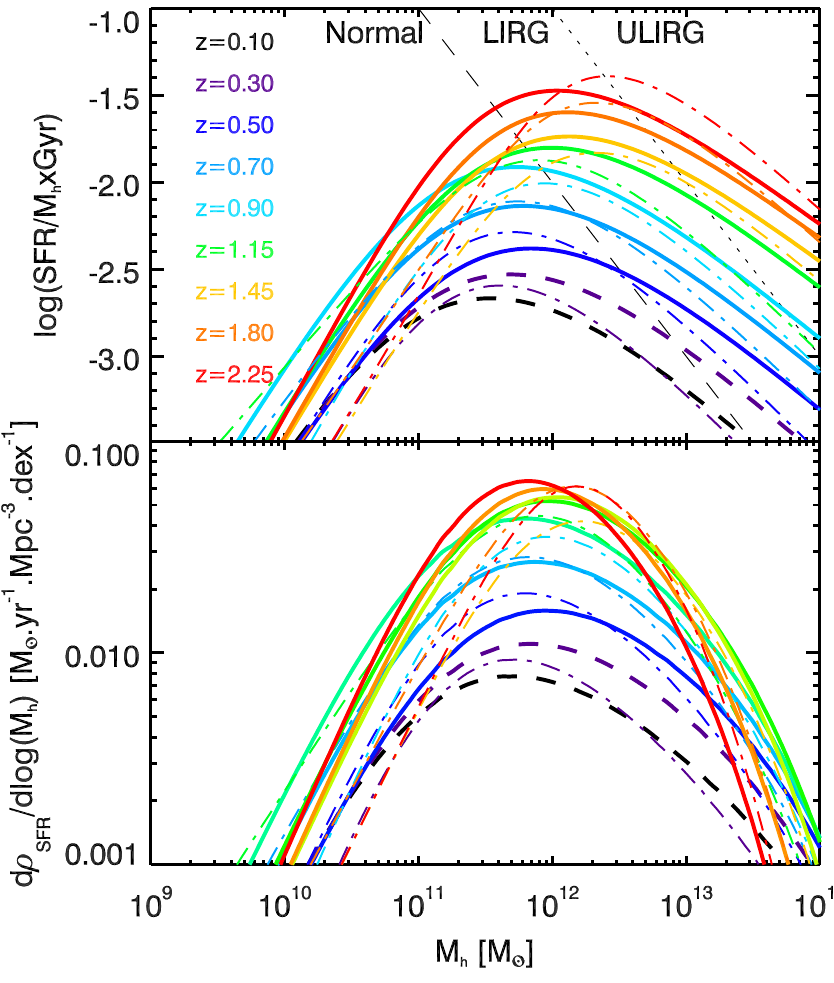}
\caption{\label{fig:ssfrhalo} \textit{Upper panel}: Ratio between SFR and halo mass as a function of halo mass at various redshifts. \textit{Solid lines}: results from a direct abundance matching of the LFIR and the HMF (we use \textit{dashed lines} at $z \le 0.3$ where neglecting the quiescent galaxies has a significant impact, see Sect.~\ref{discussion}). \textit{Dotted-dashed lines}: results from the abundance matching between the SMF and the HMF, combined with the best-fit of the sSFR-M$_{\star}$ relation by \citet{Karim2011}. \textit{Thin-dashed} and \textit{dotted} lines are the limits between the normal and LIRG, and the LIRG and ULIRG regime, respectively. \textit{Lower panel}: contribution of the various halo masses to the SFR density.}
\end{figure}

We have seen in Sect.~\ref{sm_sh} and \ref{sfr_sm} that matching abundances leads to elucidating connections between 
the stellar and halo mass on the one hand, and the stellar mass and the sSFR on the other hand. It is therefore natural to investigate the connection between the halo mass and the SFR. We can make this connection in two ways. In Fig.~\ref{fig:ssfrhalo} (upper panel), we plot the redshift evolution of the ratio between SFR and halo mass as a function of halo mass. The solid lines use a direct matching between the LFIR and the HMF. The dotted-dashed lines stem from a combination between the function linking the stellar mass and the halo mass, and the power-law fit of the sSFR-M$_\star$ relation of \citet{Karim2011} ($SFR(M_h) = f \left ( h(M_h) \right )$ with $f$ taken from the \citet{Karim2011} fit and $h$ from the results presented in Sect.~\ref{sm_sh}). In these two cases, we find a very strong evolution of the ratio between the SFR and the halo mass with redshift (1.2 order of magnitude between z=0 and  z=2.25). We also see an increase of the halo mass where the instantaneous star formation is the most efficient, from M$_h$=3$\times$10$^{11}$~M$_{\odot}$ at z=0 to  1$\times$10$^{12}$~M$_{\odot}$ at z=2.25. This evolution is even more pronounced with the second method.
The relationship between M$_h$ and SFR/M$_h$ is close to a log-normal distribution with a typical width $\sim$0.75 ($\sim$0.65 if we use the second method), but presents an excess at high mass (M$_h>10^{13}$\,M$_{\odot}$). This differs notably from the SFR-M$_h$ relation found by \cite{Conroy2009} at high mass (M$_h>2 \times10^{12}$\,M$_{\odot}$). However, at these high masses the conversion between their stellar mass growth and SFR is less reliable (see their Sect. 3.2). In contrast to \citet{Conroy2009} we made no assumption regarding the relation between star formation and halo growth.\\

From the SFR-M$_h$ relation, we can easily derive the differential contribution of each halo mass to the global SFR density $\rho_{SFR}$:
\begin{equation}
\frac{d \rho_{SFR}}{d log(M_h)} = \frac{d^2 N}{d M_h d V} \times SFR(M_h),
\end{equation}
where $d^2 N/(d M_h d V)$ is the HMF and $SFR(M_h)$ is the SFR associated to a given halo mass provided by the abundance matching. Fig.~\ref{fig:ssfrhalo} shows the evolution of this observable with halo mass and redshift. We find that the SFR density is dominated by halo masses close to $\sim$7$\times$10$^{11}$~M$_\odot$, in agreement with the analysis of \citet{Conroy2009} (their Fig.10). The distribution is quite sharp with a rapid decrease at smaller and higher mass.  We see specifically from this figure that the bulk of star formation at z$\le$2.3 never occurs in small systems. 

\section{Discussion}
\label{discussion}

We presented a first application of the abundance matching technique to current infrared data. Our exercise turned out to be surprisingly elucidating. Despite its simplicity, this technique provides a reasonable picture of the evolution of the SFR with the stellar mass, in qualitative agreement with the multi-wavelength measurements at z$>$0.3 (Sect.~\ref{sfr_sm}). It confirms the strong link between these two quantities. The success of this analysis relies on the hypothesis that the contribution of main-sequence outsider objects to the IR energy output is negligible. Matching the LFIR and the SMF including the starburst contribution into LFIR leads to a very strong increase of the sSFR-M$_{\star}$ relation for M$_{\star}>10^{11}$ M$_{\odot}$, which is not observed. Similarly, it artificially increases the sSFR-M$_{h}$ relation for M$_h>3\times10^{12}$~M$_{\odot}$. The tightness of the sSFR-M$_{\star}$ relation indicates that the SFR is not driven by merger-induced starbursts but instead by a continuous mass-dependent process that is gradually declining with time. Reproducing the evolution of this relation is a challenge for models of galaxy formation \citep[e.g.][]{Dave2008,Damen2009}. 

Our analysis neglects the presence of quiescent massive galaxies at high mass. By comparing the matched abundances with or without quiescent galaxies in the SMF using the \citet{Ilbert2010} measurements, we found that their contribution is negligible ($\le 0.2$\,dex) for $z\ge0.3$, but reach 0.25\,dex at $z=0.3$. A smaller effect ($\sim$0.15\,dex) at low $z$ is found using the measurements of \citet{VanDerWel2010}. Because the CIB is dominated by $z \ge 0.3$ galaxies, neglecting the contribution of quiescent galaxies has little impact on its modeling.

Fig. \ref{fig:ssfrhalo} demonstrates that the halo mass with the most efficient star formation does not evolve strongly with redshift, with a decrease of only a factor $\sim$3 since z$\simeq$2.3. This may contradict with one meaning of the {\it downsizing} phenomenon, which implies which stars are being formed in preferentially smaller systems at later times. This lack of evidence has also been observed by \citet{Conroy2009} up to z$\sim$1. This effect could be produced by a poor estimate of the faint-end slope of the LFIR. To investigate this, we repeated our analysis using higher slopes (up to $\alpha$=1.8) and found that our conclusion still holds. 

The existence of a characteristic mass-scale at high-mass (M$_h \simeq10^{12}$ ~M$_{\odot}$) has been known since the late 1970s and early 1980s \citep[e.g.][]{Rees1977,Silk1977,White1978}. More recently,  cosmological simulations also suggest the existence of a critical halo mass, with an important role in shaping up the main properties of galaxy. Below this mass, galaxies are built by cold flows associated with efficient star formation. Above this mass, cooling and star formation are shut down by shock-heating triggered feedback \citep[e.g.][]{Dekel2006}. Our abundance matching naturally finds this high-mass cutoff for the star-formation (although it is quite a smooth transition). At the low-mass end, there is also strong evidence for a drop in the efficiency of galaxy formation \citep[e.g.][]{Shankar2006,VanDenBosch2007,Baldry2008,Kravtsov2010,Moster2010,Guo2010,Bouche2010}. With a very crude approximation of a suppression of accretion and SFR in halo with M$_h<10^{11}$ ~M$_{\odot}$, \citet{Bouche2010} succeeded in reproducing the mass and redshift dependences of the SFR-$M_\star$ and Tully-Fisher relations from z$\sim$2 to the present. The exact physical processes leading to this mass floor are still under debate. Again, with our simple abundance matching approach, the drop of SFR in low-mass halos appears naturally. These characteristic mass-scales also roughly agree with those required by the constraints coming from the clustering modeling of star forming galaxies (e.g. \citet{Cooray2010, Planck_CIB}) although the detailed relation between our characteristic masses and theirs require additional investigations.

Because the halo mass is the prime quantity that defines the matter clustering on large scales, the link between SFR and halo mass is not only fundamental for understanding the evolution of the galaxies, but also the CIB anisotropies. The current CIB anisotropy models, based on halo occupation distribution prescriptions \citep{Penin2011,Planck_CIB,Amblard2011}, do not take into account that the brighter infrared sources tend to be in more massive halos (although \citealt{Shang2011} attempted to address this). Our analysis showed that the relation between SFR and the halo mass has a fairly constant shape from z=0 to z=2, but significantly shifts with the redshift, the characteristic halo mass varying by a factor $\sim$3 with redshift (see Fig. \ref{fig:ssfrhalo}). With improved models of CIB anisotropies, we can hope to accurately constrain this relation and its evolution with halo mass and redshift. This modeling may be able to decrease the tension between wavelengths discussed in the \citet{Planck_CIB}.

\begin{acknowledgements}
We thank Andrew Wetzel, whose questions stimulated this work, Mark Sargent, Emanuele Daddi and Lingyu Wang for their insightful comments, Gil Holder for his special perspective. MB acknowledge financial support from ERC-StG grant UPGAL 240039. Part of this research was carried out at JPL/Caltech under a contract from NASA.
\end{acknowledgements}

\bibliographystyle{aa}

\bibliography{biblio}

\end{document}